# Factors Influencing the Organizational Decision to Outsource IT Security: A Review and Research Agenda

## Full research paper

### Antra Arshad
School of Computing and Information Systems
The University of Melbourne
Parkville, Victoria, Australia
Email: atif@unimelb.edu.au

### Atif Ahmad
School of Computing and Information Systems
The University of Melbourne
Parkville, Victoria, Australia
Email: atif@unimelb.edu.au

### Sean B. Maynard
School of Computing and Information Systems
The University of Melbourne
Parkville, Victoria, Australia
Email: seanbm@unimelb.edu.au

## Abstract

IT security outsourcing is the process of contracting a third-party security service provider to perform, the full or partial IT security functions of an organization. Little is known about the factors influencing organizational decisions in outsourcing such a critical function. Our review of the research and practice literature identified several managerial factors (e.g., cost benefit, inability to cope with the threat environment) and legal factors (e.g., regulatory / legal compliance). We found research in IT security outsourcing to be immature and the focus areas not addressing the critical issues facing industry practice. We therefore present a research agenda consisting of fifteen questions to address five key gaps relating to knowledge of IT security outsourcing – i.e., effectiveness of the outcome, lived experience of the practice, the temporal dimension, multi-stakeholder perspectives, and the impact on IT security practices, particularly agility in incident response.

**Keywords:** IT security outsourcing, organizational decisions, managerial factors, legal factors, IT security outsourcing effectiveness

# 1 Introduction

A recent shift in the cyber-threat landscape has seen the emergence of organized and sophisticated actors and the use of increasingly sophisticated tools and techniques to penetrate and compromise organisations (Kotsias et al., 2022; Ahmad et al., 2019). As a result of the increasing IT security risk and the exorbitant cost of managing an IT security function, many firms are choosing to outsource IT security to a third-party provider. In fact, IT security outsourcing is on the rise and has the largest growth rate of all outsourced functions in terms of the percentage of work outsourced (Computer Economics 2021).

IT security outsourcing is the process of contracting a third-party security service provider to perform, the full or partial IT security functions of an organization (Sung and Kang 2017). The practice of IT security outsourcing can be divided into three phases, namely: the pre-outsourcing phase, the actual-outsourcing phase and the post-outsourcing phase. The IT security outsourcing ecosystem involves six major stakeholders namely: the outsourcing organization (client), the managed security service provider (MSSP), the regulatory/ governing bodies, the attacker/hacker, the insurance companies, and the organisations employees (Cezar et al. 2010; Naicker and Mafaiti 2018; Wu et al. 2021).

For organisations, the reasons to outsource IT security and the risks involved in IT security outsourcing practice make the outsourcing decision critical and challenging. Our review of the literature shows the precise factors considered by organizations in making such a decision have not been identified. Therefore, this literature review poses the following question:

*What factors influence the organizational decision to outsource IT security?*

By identifying said factors, the paper aims to provide guidance on the critical decision of IT security outsourcing to organizational decision-makers.

This paper is structured as follows. The subsequent section provides information regarding the methodology adopted for searching the relevant literature. This is followed by a thematic review of the literature. We categorize IT security outsourcing factors into two: managerial factors and legal factors. Then an analysis of the literature review is presented followed by a discussion of the gaps derived from the systematic analysis along with recommendations for future work.

# 2 Research Method

We adapted a literature review approach based on Mathiassen et al. (2007) to look at two sets of research on IT security outsourcing: the researcher perspective (using Google Scholar to identify research literature), and the practitioner perspective (using a google to identify grey literature and white papers). A four-step approach to the review was undertaken (see Table 1).

Step 1 was to identify and apply the search terms to both of our literature streams which identified 297 IT outsourcing research papers and over 17,000 hits in google looking for practitioner literature. Step 2 was to limit our search to only the top 100 papers as we found that the relevance to the topic was drastically reduced in both searches after that point. We then reviewed the papers (reading the abstracts, introductions and conclusions) from both literature streams looking to ensure that the focus was on IT security outsourcing with an additional criterion for the practitioner literature being that it had a practitioner perspective on the problem area. This resulted in 20 research and 17 practitioner articles. Step 3 was to limit the publications to "authoritative" publications – papers in good venues for research articles and for practitioner articles limiting by age (last 5 years). This resulted in 20 research and 6 practitioner articles. Step 4 (the final step) simply combined the results of the searchers resulting in 26 articles that we then carefully read and analysed using the open, axial, and selective thematic coding steps (Neuman 2014).

| Selection Step | Research literature (Google Scholar) | Practitioner literature (Google) |
|---|---|---|
| Step 1: Apply search string to search engine | **Search String:** "IT Security Outsourcing" \| "Information Security Outsourcing" - Search produced 297 results. | **Search String:** "IT Security Outsourcing or Information Security Outsourcing" report. Search produced 17,700 results. |
| Step 2: Selecting most relevant articles. | Initial 100 articles were reviewed. **Criteria 1:** The main focus is outsourcing of IT/information security. Review for relevancy determined by reviewing: abstract, introduction and conclusion of the article. Results: 20 remaining articles | Initial 100 articles. **Criteria 1**: Result illustrates practitioner perspective only. **Criteria 2:** The main focus is outsourcing of IT/information security. **Review for relevancy determined by reviewing:** title, executive summary and collected data. Result: 17 remaining articles |
| Step3: Selecting authoritative venue publications | Criteria 1: Peer reviewed articles Result: 20 | Criteria 1: Most recent publications (last 5 years). Result: 6 |
| Step 4: Combining Results | Researcher stream + practitioner stream = number of reviewed articles - 20+ 6 = 26 | |

*Table 1: Literature Search Process*

## 3  Literature Review

This section provides the outcomes of our analysis of the literature on IT Security Outsourcing. It is clear from the literature as a result of organisations relying on internet connectivity to conduct business along with increasing regulatory requirements that IT security is a critical business function with complexities and challenges (Cezar et al. 2010). However, many organizations lack the expertise or do not possess adequate resources to perform this critical security function internally (Karyda et al. 2006). Therefore, outsourcing the IT security function is an appealing option for many organizations who want to achieve higher level of security maturity (Oladapo, et al. 2009). IT security outsourcing is referred to as the establishment of a contractual relationship between organizations and sources that are external to the organization, for undertaking responsibility of its one or more IT security functions (Wilde, et al. 2006). Our review of the literature reveals that there are two categories of factors that influence an organization's decision to outsource IT security - managerial and legal. These are discussed in detail below.

### 3.1  Managerial Factors

There are eight main managerial factors that influence IT security outsourcing.

#### 3.1.1  The need for state-of-the-art expertise and resources to provide IT security

Lack of expertise and resources drives organizations toward IT security outsourcing. Effective security management requires knowledge and skills to address the diversity of security threats and escalating rates of security incidents (Karyda et al. 2006). Effective security management requires highly skilled employees, which are difficult to acquire and retain. This discourages in-house management of IT security functions and directs firms toward outsourcing their IT security. Furthermore, MSSPs tend to specialize in providing IT security services for specific industries (Cezar et al. 2017). As a result, MSSPs are better equipped in resources and expertise than their client organizations. Hence, the constraints of time, resources and expertise drives organizations to outsource their IT security (Cezar et al. 2017).

Industry literature reflects the trend towards outsourcing IT security. Deloitte's (2019) survey found the primary reasons for IT security outsourcing were: difficulty in attracting and retaining IT security experts in the competitive market, the lack of in-house technical resources required to do cyber security right. PwC (2017) estimates that the unfilled cybersecurity positions worldwide would be an astounding 3.5 million, which will drive more organizations to outsource their IT security function.

### 3.1.2 Cost savings

Literature advocates that there are cost savings associated with the outsourcing of IT security for the client organization. Therefore, the advantage of cost saving favours outsourcing of IT security. Research studies infer that many firms decide to outsource security operations because cost is the principal concern and there are benefits in the form of cost savings or security improvements per dollar of investment (Naicker and Mafaiti 2019). A critical factor in outsourcing decisions is the relative cost efficiency of MSSPs (Hui, et al. 2012; Lee, et al. 2013; Wu et al. 2020). However, our review of these studies shows that the risk associated with outsourcing to a third party was not adequately considered.

### 3.1.3 Internal security culture

Outsourcing IT security has negative consequences for the overall security culture of the organization. Security culture is weakened when IT security outsourcing is practiced (Dhillon et al. 2017; Sung and Kang 2017, Oladapo et al. 2009). Further, removing the internal security may send a message that security is not important. Outsourcing IT security diminishes security awareness and sense of responsibility among the employees of the client organization (Karyda et al. 2006).

Deloitte (2019) also reports that "in this age of digital transformation, perception leads reality when it comes to organizations' ability to create a cyber-secure culture". Executives will need to frequently reconsider their organizations' security strategies for a stronger internal security culture, through close monitoring of the evolving ecosystem of specialized contractors and outsourcing providers available to help manage cyber risk (Deloitte 2019).

### 3.1.4 Threat environment

Several surveys conducted in the last few years conclude that besides hacker's behaviour, high rates of cyber-attacks and constant high-profile security threats compel organizations to outsource their IT security (Info-Security Europe 2015; Avasant 2020; Syntax 2021) as there is a feeling that they are unable to cope with the constant attack pressure from the threat environment.

However, the research literature presents contrasting arguments. A cyber-threat actor's behaviour and frequency of attack should be considered by organizations before making an outsourcing decision (Hui et al. 2012; Syntax 2021; Wu, et al. 2020). Wu et al. (2020) points out that "a strategic hacker always has a higher incentive to attack firms that manage security in-house, which reveals the advantages of outsourcing the security service to an MSSP". Whereas other studies argue that cyber-attacks may be more massive and pervasive when hackers are able to breach MSSP's adopted security technologies, as multiple clients adopt similar security technologies provided by the MSSP (e.g., Hui et al. 2012).

### 3.1.5 Characteristics of MSSP

Literature discusses the characteristics of MSSPs (such as their cost efficiency, their technical advantage, and how compatible legally with the client they are) as the merits that organizations should look for while finalizing the security outsourcing agreement (Hui et al. 2012; Karyda et al. 2006; Lee et al. 2013; Wu et al. 2017). "Competence of the vendor to ensure information security" and "compliance of the vendor with client requirements and external regulations" are two important characteristics of MSSPs inferred in literature (Dhillon et al. 2017).

Further, the prospects of IT security outsourcing increases if the MSSP has a partnership with a technology vendor, as muti-sourcing IT security contracts that involve both service providers and technology vendors, benefit the client organization (Naicker and Mafaiti 2019). "For the client, access to the service provider brought access to the technical skills from the vendor as well" (Naicker and Mafaiti 2019).

### 3.1.6 IT value proposition

Researchers believe that organizations should consider their IT value proposition before outsourcing their IT security functions (Fenn et al. 2002; Feng et al. 2019; Karyda et al. 2006). Literature argues that organizations that regard IS/IT security as a commodity asset are more likely to outsource their security functions to concentrate their resources on core competencies (Fenn et al. 2002; Karyda et al. 2006). The literature also argues that in those firms where IS/IT provides a strategic advantage thus having IT closely connected to their core competencies, they tend to keep their IT security function in-house (Feng et al. 2019). However, the practice literature shows that most organizations undergoing digital transformation tend to outsource their IT security functions (Syntax 2021).

### 3.1.7 Outsourcing contract structure

The structure of an IT outsourcing contract can mitigate client organizations' concerns with outsourcing IT security. For example, a multilateral contract structure can solve issues of the double moral hazard, regardless of the externality (Cezar et al. 2014; Lee et al. 2013). Similarly, researchers advocate penalty-based contracts consisting of fixed service fees and penalty for degraded services (Cezar et al. 2010; Feng et al. 2019; Hui et al. 2012). Moreover, inclusion of contractual warranties that ensures MSSP compliance with the client's policy, favourably effects an organization's decision to outsource. For example, IBM's Internet Security System SLA contains a $50,000 money-back warranty for each breach offered to win security outsourcing clients (Fenn et al. 2002; Lee et al. 2013). PwC (2017) illustrates that inclusion of threat intelligence sharing in outsourcing contracts results in better prospects of outsourcing IT security to managed service providers.

### 3.1.8 Outsourcing organization's IT governance structure

IT governance structures should be considered by organizations before making IT outsourcing decisions, as it has effects on the results of outsourcing decisions (Liu et al. 2017). Their study illustrated that educational institutions with centralized IT governance structures benefit more from outsourcing their IT security as compared to institutions with decentralized IT governance. Centralized IT governance impacts the regulatory compliance of an organization which ultimately impacts the occurrence of cyber security breaches. Due to this better compliance and system integration, these institutions have coordinated responses to security events and can utilize security outsourcing for further lowering the probability of encountering cybersecurity breaches (Liu et al. 2017). PwC (2017) also shows that organizations with a centralised governance model tended to benefit more from outsourcing IT security as they can have centralised threat management and response to sharing of threat intelligence.

## 3.2 Legal Factors

There are three key legal factors that influence IT security outsourcing. These are discussed below.

### 3.2.1 Regulatory / legislative compliance

Researchers have identified that evolving regulatory requirements trigger the security outsourcing practices (Cezar et al. 2014; Naicker and Mafaiti 2019). The need for new regulations is due to evolving technological trends. Since compliance with regulatory requirements demand resources and expertise, therefore firms outsource to transfer the burden of this ever-changing mandate. Literature categorises compliance with legislation in the very high impact category as "when the trigger to outsource security services is required by legislation; outsourcing is highly likely" (Oladapo et al. 2009). In industry practice, regulatory compliance is one of the reasons CISOs prefer to outsource IT security (Bissell et al. 2020). Due to the increase in indirect attacks, organizations also need to secure their third parties and partners. However, regulatory compliance is a major challenge in managing third party cyber risks due to third party organisations being in different legal jurisdictions (and countries). Therefore, organizations tackle this wider scope and scale of IT security by outsourcing it to managed security service providers (Bissell et al. 2020).

### 3.2.2 Data protection requirements

While deciding in favour of IT security outsourcing, organizations should ensure that they are not violating their commitment to data protection laws, especially if the outsourced function includes personal data (Cezar et al. 2017; Fenn et al. 2002; Hui et al. 2012; Karyda et al. 2006). Breaches of data privacy and protection is also reported as one of the main risks associated with IT security outsourcing (Bissell et al. 2020; Ernst & Young Global 2019). Interestingly, "fines excess of US$100 million resulting due to violations of general data protection regulations (GDPR), may match, or even exceed, the overall cost of cybercrime for an organization" (Bissell et al. 2020). Therefore, organisations are sceptical for outsourcing IT security due to data protection requiemetns.

### 3.2.3 Protection of intellectual property

Although there is little discussion in literature regarding legal implications involving intellectual property in security outsourcing, it is a critical factor that organizations consider while making outsourcing decisions. The protection of intellectual property generates the need to acquire licences to retain rights while entering into a security outsourcing contract (Fenn et al. 2002). Therefore, the involvement of intellectual property in security outsourcing can offset the financial viability of the outsourcing decision (Dhillon et al. 2017). Moreover, a lack of compatibility between the applicable law of the client and the MSSP, under which the intellectual property is protected may result in loss, misuse and damage to intellectual property (Karyda et al. 2006). Therefore, protection of intellectual property becomes an important issue when considering IT security outsourcing.

### 3.3 Summary

There are eight managerial and three legal factors that can favour or hinder an organizational decision to outsource its IT security function. An organization can make an informed decision regarding its IT security outsourcing by carefully considering the factors to achieve IT security goals through outsourcing. The next section discusses the findings from the literature review.

## 4 Discussion

There is very limited research literature available in the field of IT security outsourcing despite its industry wide adoption. The conversation in literature is mainly dominated by the discussion on the considerations of the pre-outsourcing phase and from the outsourcing organization's perspective. The literature emphasizes: the conditions in which organizations choose/should choose to outsource IT security; complete or partial IT security outsourcing practices; the features or characteristics of MSSP that are desirable in terms of outsourcing IT security; and IT security contract finalization.

Research in IT security outsourcing is immature and does not address the major areas of potential opportunity and issues which ultimately results in research gaps. As mentioned previously, IT security outsourcing is the process of contracting a third-party security service provider to perform the full or partial IT security functions of an organization. It has three temporal phases (pre-outsourcing, actual outsourcing and post-outsourcing) and has six major stakeholders. Based on the review and given the definition of IT security outsourcing, we present a research agenda based on four gaps in knowledge.

### 4.1 Gap 1: Effectiveness of IT Security Outsourcing is Not Addressed

There is no discussion in literature regarding the success of IT security outsourcing in producing the desired results. Therefore, how can the success of IT security outsourcing be measured? This potential area of opportunity could be addressed through studies of success factors, maturity models, cost benefit tools, and strategic evaluations of IT security outsourcing decisions. The effectiveness of IT security outsourcing is a key area of interest for many practitioners which could have been addressed through studies of the post outsourcing phase. Table 2 shows that little research is conducted around activities (including effectiveness) which would occur in the Post Outsourcing phase.

| Outsourcing Phase | IT Security Outsourcing Literature |
|---|---|
| Pre-Outsourcing | Cezar et al. (2010); Cezar et al. (2014); Cezar et al. (2017); Dhillon et al. (2017); Feng et al. (2019); Fenn et al. (2002); Hui et al. (2012); Hui et al. (2019); Karyda et al. (2006); Lee et al. (2013); Naicker and Mafaiti (2019); Oladapo et al. (2009); Samarasinghe et al. (2007); Sung and Kung (2017); Wu et al. (2021); Wilde et al. (2006); Zhang et al. (2021). |
| Actual Outsourcing | Cezar et al. (2014); Hui et al. (2012); Hui et al. (2019); Sung and Kung (2017). |
| Post Outsourcing | Rowe (2008). |

*Table 2: Distribution of literature studies by the phase of outsourcing*

### 4.2 Gap 2: The Lived Experience of IT Security Outsourcing has not been Addressed

The following critical issues and variables were not considered in extant studies: organizational profile (in terms of size, scale and industrial sector), risk profile, and the organizational context. This suggests that the approach taken by researchers toward IT security outsourcing has not considered the lived experience (the organizational context) within which IT security outsourcing is conducted. One suspected reason for this gap, as shown in table 3 could be that the scholars contributing to literature regarding IT security outsourcing do not have an IT security background. An informal review of the background of researchers shows the majority do not have a track record of conducting IT security research and practice – hence a full understanding of the practice of IT security is missing from their interpretations. Rather they tend to discuss IT security from a socio-economic perspective illustrating its economic considerations and social impacts.

| Literature Stream | IT Security Outsourcing Literature |
|---|---|
| Information Security | Fenn et al. (2002); Oladapo et al. (2009); Samarasinghe et al. (2007), Wu et al. (2021). |
| Economics of Information Security | Cezar et al. (2010); Cezar et al. (2014); Cezar et al. (2017); Feng et al. (2019); Lee et al. (2013); Liu et al. (2017); Zhang et al. (2021). |
| Information Security & Sociology | Dhillon et al. (2017); Hui et al. (2012); Hui et al. (2019); Karyda et al. (2006); Naicker and Mafaiti (2019); Sung and Kung (2017); Wilde et al. (2006). |

*Table 3: Distribution of literature studies by theme*

## 4.3 Gap 3: Temporal Dimension of IT Security Outsourcing is Not Explored

IT security outsourcing changes over time. Initially, in-sourcing was the predominant outsourcing technique, then it was outsourcing and now the shift is moving back to insourcing or part-sourcing. These temporal dimensions of IT security outsourcing are not addressed within the IT security outsourcing literature. Therefore, how often researchers need to revisit the question of studying outsourcing remains unclear. The temporal dimensions of IT security outsourcing ensures that: IT security outsourcing is discussed as a phenomenon in research and its researchers are cognisant of the changes in the climate around IT security outsourcing in industry. Moreover, it ensures that researchers are keeping pace with the changing industry and market trends. So, there is a need for alignment between changing trends in industry and the studies done by the researchers to understand the IT security outsourcing phenomenon and how it is evolving.

## 4.4 Gap 4: A Multi-Stakeholder Perspective of IT Security Outsourcing is Not Being Addressed

The role of stakeholders is key to understanding the holistic ecosystem however the literature of IT security outsourcing focuses on the organizations taking the decision to outsource. The literature has either completely ignored primary stakeholders (e.g. insurance companies and the staff belonging to the outsourced function), or it discusses stakeholders (e.g. MSSP, hackers and Governing bodies) in relation to the interest of the outsourcing company. This has resulted in the dominance of a single stakeholder perspective in literature discussion. Moreover, the holistic view of the IT security outsourcing ecosystem may have revealed indirect reasons to outsource IT security like gaining market repute and utilizing a big brand name to win customer trust rather than functional and operational reasons for being secure.

## 4.5 Gap 5: The Impact of IT Security Outsourcing on IT security practices is Not Considered

There has been considerable research on identifying IT security practices of organizations (e.g., risk management, policy, training) (Alshaikh et al., 2014; Maynard et al., 2011). The impact of outsourcing IT security on said practices have not been studied. A particularly interesting case is that of incident response (Kotsias et al., 2022; Ahmad et al., 2021). The performance of an IT security incident response is measured in terms of agility (swiftness, flexibility, innovation) which is particularly challenging given MSSPs tend to address IT security from a technology maintenance perspective rather than a business strategy perspective (Naseer et al., 2021).

## 4.6 Research Questions to Address Gaps in Knowledge

From all the gaps identified and discussed above it can be concluded that the current state of research lacks maturity. Table 4 provides the summary of the gaps identified and the potential research questions that can be answered for addressing these gaps.

| Sec | Gap Identified | Potential Research Questions |
|---|---|---|
| 4.1 | The effectiveness of IT security outsourcing is not addressed | • What are IT security outsourcing success factors?<br>• What is maturity in IT security outsourcing?<br>• What is the strategic value of IT security outsourcing? |
| 4.2 | The lived experience of IT security outsourcing has not been addressed | • How is the effectiveness of outsourcing IT security influenced by organizational risk, organizational context and the threat environment?<br>• What are the effects of IT security outsourcing on internal security culture and in-house security functions of an outsourcing organization?<br>• How often should IT security outsourcing contracts be reviewed (given the complex and evolving security landscape and organizational context)?<br>• How can an outsourced IT security function be audited (considering the multiple stakeholder perspective)? |
| 4.3 | Literature does not consider the temporal dimension of IT security outsourcing | • How has IT security outsourcing evolved over time (given industry trends, norms and expectations)?<br>• What is the current state of IT security outsourcing within the broader landcape of evolving trends? |
| 4.4 | A multi-stakeholder perspective is not being addressed | • What is the role of stakeholders in IT security outsourcing (including insurance providers, regulators, and government)?<br>• What expectations and requirements are imposed by parties involved in IT security outsourcing contracts?<br>• How do MSSPs identify and assess their client's security requirements as part of operationalizing an IT security outsourcing contract?<br>• What are the drivers (direct or indirect) for IT security outsourcing (e.g. contracting reputed MSSPs to improve shareholder confidence)? |
| 4.5 | The Impact of IT Security Outsourcing on IT security practices is Not Considered | • What is the impact of IT security outsourcing on IT security practices in organizations (e.g. strategy, policy, training, risk management)?<br>• What is the impact of IT security outsourcing on the agility of incident response? How should command and coordination be structured and instituted across organizational boundaries? |

*Table 4: Summary of gaps identified and the potential research questions.*

## 5  Conclusion

This paper conducted a literature review and developed a research agenda on the factors influencing organizational decisions to outsource IT security. We found two categories of factors that influence an organization's decision to outsource IT security - managerial and legal. The managerial factors were: (i) The need for state-of-the-art expertise and resources to provide IT security, (ii) Cost benefit, (iii) Internal security culture, (iv) Threat environment, (v) Characteristics of MSSP, (vi) IT value proposition, (vii) Outsourcing contract structure and (viii) Outsourcing organization's IT governance structure. The legal factors were: (i) Regulatory/ legislative compliance, (ii) Data protection requirements and (iii) Protection of Intellectual Property.

Our analysis of the literature led to the development of a research agenda based on five gaps in knowledge. We suggest future research into IT security outsourcing study effectiveness, lived experience, temporal dimensions, multi-stakeholder perspectives, and the impact on IT security practice, particularly agility of incident response.  We have proposed a list of research questions for each gap in knowledge. We believe that research addressing these questions will make substantive contributions to both theory and practice.

Finally, we articulate the key insights from this paper. We found that industry practice of IT security outsourcing is primarily driven by financial considerations without due consideration of IT security risk.

We found literature arguing the risk of information leakage/IP can offset cost savings made by outsourcing IT security. Further, there are considerable advantages in terms of cost efficiency, technical advantage and legal compatibility in outsourcing IT security where the outsourced entity is a partnership between an MSSP and a technology vendor. Adopting a centralised IT governance structure when outsourcing IT security yields improved compliance and resource management benefits. Finally, that evolving regulatory requirements is a reason why organizations outsource their IT security.